# X-ray photoemission spectrum, electronic structure, and magnetism of $UCu_2Si_2$

[1]J.A. Morkowski, [2]G. Chełkowska, [1]M. Werwiński, [1,*]A. Szajek, [3]R. Troć, [4]C. Neise

[1] Institute of Molecular Physics, Polish Academy of Sciences
M. Smoluchowskiego 17, 60-179 Poznań, Poland

[2] A. Chełkowski Institute of Physics, Silesian University
Uniwersytecka 4, 40-007 Katowice, Poland

[3] W. Trzebiatowski Institute of Low Temperature and Structure Research
Polish Academy of Sciences
P.O. Box 1410, 50-950 Wrocław, Poland

[4] IFW Dresden, P.O. Box 270116, D-01171 Dresden, Germany

**Abstract**

The room temperature X-ray photoemission spectrum of the ferromagnetic compound $UCu_2Si_2$ ($T_C$ = 100 K) was measured using an Al $K_\alpha$ source. Related theoretical spectra were computed from densities of electronic states obtained in the local density approximation (LDA), the generalized gradient approximation (GGA), and using the GGA+$U$ method. The calculated spectrum is in a good agreement with the experimental one. The spin polarized calculations based on the GGA+$U$ approach as well as GGA/LSDA with orbital polarization (OP) corrections taken into account provide values of the total magnetic moment in reasonable agreement with the experimental values ranging between 1.6 and 2.0 $\mu_B$/U atom.

[*]Corresponding author: email address: szajek@ifmpan.poznan.pl

# 1. Introduction

The rare earth (R) and actinide elements (An) form a large family of intermetallic compounds of general chemical formula $(R,An)T_2(Si;Ge)_2$ (T = transition metal), crystallizing in the bct $ThCu_2Si_2$ structure type. Many of them exhibit interesting physical states, as for example heavy fermion, anisotropic superconducting, or different types of magnetically ordered states [1]. The most puzzling member of this group seems to be $URu_2Si_2$ that shows a coexistence of superconductivity and so-called hidden order [1].

Among the ternary uranium silicides $UT_2Si_2$, the copper- and manganese-containing ternaries are only ferromagnets with $T_c$ = 103(2) and 377 K, respectively. The remaining members of this family of ternaries are either Pauli paramagnets (T= Fe and Os) or simple (T = Cr, Co, Rh and Ir) and more complex multiphase (T= Ni, Pd and Pt) antiferromagnets. As shown by powder [2] and single-crystalline [3] neutron diffraction experiments, the Cu-based silicide possesses an ordered moment of 1.6 $\mu_B$ or 2.0 (1) $\mu_B$, respectively, oriented along the [001] direction (c-axis). Moreover, earlier studies of $UCu_2Si_2$ performed on polycrystalline samples suggested that also an antiferromagnetic order appears within a few degrees above $T_C$ [4]. Quite surprisingly, the first single crystal study of $UCu_2Si_2$ obtained by the Cu-flux method has revealed an additional antiferromagnetic ordering below $T_C$ with $T_N$ = 50 K [5]. At variance to this finding, more recent single crystal studies of this compound, now using crystals grown by the Sn-flux procedure [6], have indicated that besides a ferromagnetic state stable in the whole temperature range below $T_C$ = 100 K there exists antiferromagetic order between $T_C$ and $T_N$ = 106 K [3, 6]. Neutron examination of this antiferromagnetic phase [3] revealed an incommensurate longitudinal spin-density wave state (SDW) with a long period-icity $\Lambda$ = 85.7 Å and nearly sinusoidal magnetic modulation with a propagation vector **k** = [0 0 0.116]. The magnetic structure contains as many as 17 layers with in-plane ferromagnetic order. On the other hand, another single-crystalline sample of $UCu_2Si_2$ obtained by the Cu-flux method [7] like the single crystal of Ref. [5] has not shown any antiferromagnetic phases, below or above $T_C$. The saturation magnetic moment was determined to be 1.55 $\mu_B$, a value close to that obtained by powder neutron diffraction [2]. Somewhat higher magnetization value of 1.75 and neutron value of 2.0(1) $\mu_B$-were reported in Refs. 6 an 3, respectively. However the latter high value of this moment was discussed by the authors [3] as being estimated with some large error.

A comprehensive paper on the electronic structure of $UT_2Si_2$ compounds, determined by means of self-consistent density-functional calculations in local spin density approximation (LSDA), treating the U *5f* states as band states, has been published by Sandratskii and Kübler [8]. These calculations were carried out with the fully-relativistic augmented spherical wave (ASW) method. The electronic structure of this series of compounds can be characterized by the relative energy positions of the *d* states of a given T atom compared with those of the uranium *5f* states. For both $UCu_2Si_2$ and $UPd_2Si_2$ the T *d* states lie lower than the *5f* states and are separated from the latter by an energy interval of about 4 eV and 2 eV, respectively. This interval becomes smaller with decreasing atomic number of the T atom due to the lower *d*-occupation and the related upward-shift of the *d*-band center. Finally, *d*- and *5f*-band merge and are strongly hybridized. Thus, in the case of $UCu_2Si_2$ where the *d*- and *f*-states are farthest away in energy, *f-d* hybridization is considered to be the smallest among all uranium 1:2:2 type compounds. At the same time, the highest density of states (DOS) N at the Fermi level $E_F$ among these ternaries is found in $UCu_2Si_2$. The calculated uranium moments (0.88 $\mu_B$) are much smaller than the related experimental values (between 1.6 and 2.0 $\mu_B$). A continuation of the work presented in the Ref. [8] was made by Mavromaras, Sandratskii and Kübler [9], where so-called orbital polarization (OP) corrections were applied. They found that this approach leads to better agreement between theory and experiment, in particular concerning the magnitude of the total magnetic moment of $UCu_2Si_2$ (2.0 $\mu_B$ in the LSDA+OP calculation).

In this paper we present results of three kinds of band structure calculations using different approaches, being discussed in detail below. Due to the fact of the first performed XPS measurements on $UCu_2Si_2$, using single-crystalline sample it was possible to compare the calculated data with the high quality observed photoemission intensities in the valence band energy range. As we will show below, the obtained good agreement between both the results indicate a proper treatment of this problem. The reason for using several methods of electronic structure calculations was to clarify marked discrepancies between the calculated and experimental values of magnetic moment on uranium atom.

## 2. Experimental and Computational Details

The X-ray photoemission spectroscopy (XPS) measurements were performed on single crystals of $UCu_2Si_2$ obtained by the procedure described in Ref. [7]. The XPS measurements were carried out using a PHI 5700/660 Physical Electronics Spectrometer with Al $K_\alpha$ source. The spectra were analyzed by a hemispherical analyzer with an energy resolution of about 0.3 eV. Clean samples were obtained by scraping the surfaces with a diamond file. The pressure was in the low $10^{-10}$ Torr range. All measurements repeated several times were carried out at room temperature. A standard procedure of subtracting the background based on the Tougaard method [10] was applied followed by a deconvolution of the total core level curve using the Doniach-Šunjić-type expression [11].

The band structure calculations were done by three methods: (i) the Full-Potential Local-Orbital code (FPLO) [12], (ii) the Full -Potential Linear Augmented Plane Waves (FP-LAPW) method implemented in the latest Wien2k-version of the original Wien code [13] and (iii) the full-potential LMTO method in the implementation of the LmtART code (version 6.50) [14, 15]. These methods use different sets of basic functions, different treatment of relativistic and correlation effects (see the references [12-15] and below). The reason for using several methods of electronic structure calculations was to clarify marked discrepancies between the calculated and experimental values of magnetic moment on uranium atom. As an additional test we compare experimental and calculated photoemission spectra.

In order to check the sensitivity of the results with respect to the treatment of the exchange-correlation functional, we used different versions of exchange-correlation potentials implemented in the computational methods. The calculations with the Wien2k code were performed in the generalized gradient approximation (GGA) using the parameterization proposed by Perdew, Burke and Ernzerhof [16]. The FPLO calculations were carried out in LSDA (parameterization by Perdew and Wang [17]) and the LmtART-results were obtained as well in LSDA (parameterization by Vosko-Wilk-Nussair [18] with gradient corrections included [16]). For some calculations the orbital polarization (OP) correction was taken into account [19-21] in both FPLO and Wien2k codes. Further Wien2k-calculations were carried out using the GGA+$U$ approach in the version introduced by Anisimov et al. [22] with an approximate self-interaction correction (SIC) implemented in the rotationally invariant way according to Liechtenstein et al. [23]. A detailed discussion on expression for total energy and the double counting term specification in GGA+$U$ methods implemented in the Wien2k code is given in [24]. Values of the *5f* on-site Coulomb energy $U$ in the range from 0 to 6 eV and exchange parameters *J* between 0 and 0.5 eV were tested to get the correct value of the uranium magnetic moment as well as to reproduce the experimental photoemission spectrum. The calculations were performed without and with spin polarization, respectively, to calculate photoemission spectra and to calculate magnetic moments.

Relativistic effects were taken into account with appropriate attention: in the LmtART and Wien2k codes, spin-orbit interaction was treated in the so-called second-variation approach on top of a scalar relativistic calculation, while the FPLO calculations were performed in full-relativistic mode.

Room-temperature (RT) lattice parameters, a = 0.3985 nm, c = 0.9945 nm were employed in all calculations [7]. The atomic positions in the unit cell are: U(2b) = (0, 0, 1/2), Cu(4d) = (0, 1/2, 1/4), Si(4e) = (1/2, 1/2, 0.6177) [7].

The FPLO basis set comprised the following *(semicore:valence)* states: Cu *(3s3p:3d4s4p)*, U *(5d6s:5f6d7s7p)*, Si *(2s2p:3s3p3d)*. For the Wien2k calculations, $RK_{max} = 8$ was chosen and the muffin-tin radii were equal to 2.50, 2.31, and 2.04 a.u. for U, Cu, and Si atoms, respectively.

A tetrahedron method [25] was employed to carry out the k-space integrations, using at least 8000 points in the Brillouin zone (BZ), i.e. 828 points in the irreducible wedge of the BZ. The total energy was converged at a level of $10^{-6}$ Ry.

In order to calculate the X-ray photoemission (PE) spectrum, the partial site contributions to the density of states in the valence band region were multiplied by the corresponding PE cross sections (tabulated in Ref. [26]) and convoluted with a Gaussian profile of 0.3 eV width. As the calculated photoemission spectrum relies on tabulated [26] values of the photoemission cross sections, calculated up to 2 significant digits (for partial cross sections used in the present paper), the overall accuracy of the photoemission intensity is no better than a few percent. However, only relative intensities of the features/peaks of the photoemission spectrum will be affected by uncertainties of the partial cross sections used in the calculations, not their positions on the energy scale.

## 3. Results and Discussion

The obtained XPS results are presented in Fig.1. Figure1(a) shows the valence band photoemission spectrum of $UCu_2Si_2$ in an energy range of 38 eV. There are two prominent features which correspond mainly to the U *5f* states at about 0.3 eV below the Fermi level ($E_F$, at energy zero) and to the Cu *3d* states at about 4.3 eV below $E_F$, respectively. The above spectrum includes also the U *6p* states split by spin-orbit (S-O) interaction. In Figure 1(b) we have plotted the related spectrum of the U *4f* core levels. The characteristic fine structure of these peaks provides an interesting indirect information on the *5f* states in this silicide, which will be discussed below. The U *4f* spectrum shows a spin-orbit splitting of 10.9 eV between the *4f$_{7/2}$* and *4f$_{5/2}$* components, each of them consisting of a dominant asymmetric main line and three satellites on the higher binding energy (BE) side. This kind of multi-satellite structure has been observed in the core-level spectra of other uranium compounds studied by us on single-crystalline samples, not being contaminated by oxygen like we will specify below: $U_3M_2M'_3$, M = Al, Ga, M' = Si, Ge [27], UAuSb2 [28], UN [29], $UGe_2$ [30].

The peaks of the U $4f_{7/2}$ and U $4f_{5/2}$ main lines are situated at 376.8 eV and 387.7 eV, respectively. The singularity index α in the Doniach-Šunjić-type expression [11] was assumed in the fitting procedure to be about 0.5. The presence of the additional final state satellites: sat. I, sat. II, and sat. III at about 1.3, 3.7, and 7 eV higher BE with respect to the main lines may be interpreted as resulting from a contribution of *$5f^2$*, *$5f^3$* and *$5f^4$* final states, respectively [31]. The satellite peak at about 380.5 eV, i.e. sat. II of the *$4f_{7/2}$* component, might be enhanced by oxidization of the sample surface, resulting in $UO_2$ with a peak at a similar position [32]. It is well known that uranium compounds are very reactive and may oxidize even under high vacuum conditions. On the other hand, the amount of oxygen directly detected in the sample was completely negligible (see the O *1s* position in Fig. 1(c)). Unfortunately, it was impossible to make such a check in the valence region spectrum near 5.5 eV BE, since the strong Cu *3d* emission covered that position. The features at 384 eV and 394.9 eV (satellites III) are often referred to as the 7 eV satellites. The presence of this kind of satellites, occurring in the *4f* photoemission spectrum of many uranium intermetallic compounds, is regarded as an indication for localization of the *5f* electrons [29, 31, 33, and references therein].

In earlier XPS studies such a group of three satellites was observed, e.g., in the case of the pseudobinary solid solutions $URh_{1-x}Pd_x$ (Fujimori *et al*. [34]) and in the heavy-fermion ternary systems $UT_2Al_3$ (T ≡ Ni or Pd) (Fujimori *et al*. [35]). In turn, in our earlier studies the three-line structure of the *4f* core level spectra was also observed in $U_3M_2M'_3$, M = Al, Ga, M'= Si, Ge [27], UN [29], and most recently in $UGe_2$ [30]. As an alternative to the explanation given above, these satellites were ascribed to the dual character of *5f* electrons which seems to be common for many uranium intermetallics. This point of view has been developed theoretically by Miyake and Kuramoto [36] and was applied to the case of $UPt_3$ by Zwicknagl et al. [37]. A recent experimental confirmation for the case of the $UT_2Al_3$ type of ternaries was published by Sato *et al*. [38].

Results of band structure calculations are presented in Fig. 2, where the left hand panel shows DOS of the valence bands based on different approaches applied. Special treatments, as LSDA/GGA+OP and GGA+*U,* concern the U *5f* electrons and as this panel indicates the main differences are observed in the nearest vicinity of the Fermi level where the U *5f* states provide the main contribution to the valence band. The right hand panel presents DOS plots obtained using the Wien2k code with the GGA+*U* approach, with local contributions to the total DOS resolved. In contrast to the uranium-projected DOS, no significant differences between GGA+*U* and GGA (the latter are not shown) are detected in Cu- and Si-projected

DOS because no corrections to the GGA are applied for the Cu- and Si-states. Furthermore, in Fig. 2 one can observe some distinct features: an approximately 1 eV wide band dominated by U *5f* states just below the Fermi level; a well-defined, 2 eV wide band centered at ~ -4 eV due to the Cu *3d* states; a separate band in the BE range between roughly -11 and –7 eV dominated by the Si *3s* states; finally, the S-O split U *6p_{3/2}* and U *6p_{1/2}* bands located around – 16 eV and -23 eV, respectively. Only the U *5f* bands show a significant split due to spin polarization, while the other bands are only weakly spin polarized. Some numerical characteristics of the DOS at the Fermi level are collected in Table 1.

The calculated X-ray photoemission spectrum together with the measured one is presented in Fig. 3. The theoretical LDA/GGA curves shown in Figure 3 (a, b) were calculated with the FPLO and Wien2k methods. The LmtART method gave virtually the same plots. The difference in the calculated S-O splitting of the U *6p* states results from different treatments of relativistic effects: full-relativistic (FPLO code) vs. second-perturbation approach (Wien2k code). There is an overall agreement with experiment but some quantitative differences exist. The calculated *5f* maximum is somewhat closer to the Fermi level than the measured peak. More obvious, we notice a shift of the calculated Cu *3d* peak towards lower binding energy. Similar differences between GGA and measured peak positions were found for several systems with completely occupied *d* band and explained with incomplete screening of the final states [39].

Here, we model these effects with the GGA+*U* method [22-24]. Results of such calculations, using several choices of Coulomb *U* and exchange *J* parameters [40] are presented in Fig. 3 (c) and again compared with the experimental valence band photoemission spectrum. Although the position of the U *5f* feature is quite insensitive to the choice of the above parameters *U* and *J*, the broad peak around 4 eV BE, originating from the Cu *3d* electrons, matches the experimental position for $U_{3d}$ = 2.5...3.2 eV and $J_{3d}$ = 0 eV. However, the width (at half intensity) of the experimental photoemission peak is larger by about 0.4 eV than the calculated one, nearly independent of *U*. Similar approach was made in the case of $Ce_2CoSi_3$ [41], where different parameters *U* were used for Ce(*4f*) and Co(*3d*) electrons.

In Fig. 4 we show the dependence of the uranium-projected magnetic moments on the parameters $U_{5f}$ and $J_{5f}$ within the LSDA+*U* approach using the Wien2k code. Experimental values reported so far in the literature by different authors are marked by horizontal lines. As seen from this figure, the curves calculated for distinct *J*-values tend to saturation with increasing *U* and reach the lowest experimental value given in Refs. [2, 7] for $J_{5f}$ = 0.5 eV and $U_{5f}$ = 3 eV or above. We note a weak dependence on *U* for *U* > 2 eV, but a fairly strong dependence on *J*. Thus, the resulting magnetic state is quite sensitive to the choice of parameters. This finding leads us to advocate the LSDA+OP approach which is free of adjustable parameters. Indeed, the results obtained by using the OP corrections are distributed around the highest experimental value of 2.0 $\mu_B$ [4]. It appears that the magnetic moments calculated within the LSDA were found too small because of obtaining near cancellation of spin and orbital moments. The numerical values of spin- and orbital contributions as well as total moments are compiled in Table 2.

We finally note that the LSDA/GGA magnetic moments obtained with the full-potential codes Wien2k and FPLO almost coincide. The same holds for the spin moments obtained by LSDA/GGA+OP calculations with these two codes, while the related orbital moments differ by almost 10%. We attribute this difference to the code-dependent definition of atomic orbitals and the related differences in the Racah parameters. On the whole, the comparison in Table 2 gives an idea about the deviations between numerical results obtained with different electronic structure codes.

## 4. Conclusions

The relatively broad U *5f* band and the fair agreement between the experimental photoemission spectrum and the LDA-derived spectrum are indications of distinct delocalization of U *5f* electrons in $UCu_2Si_2$. The observed deviation in the *3d* peak position between calculated and experimental spectrum can be due to incomplete screening of the final state not included in the LDA approach. It can be modeled by using LSDA+*U* with a decent parameter value. Also, the significant differences between LSDA/GGA and experimental magnetic moments are reduced by taking into account effects of the orbital polarization. We predict a ratio of spin- vs. orbital magnetic moments of about 1/-2.

**Acknowledgments**

This work was supported by the funds for science in years 2007–2010 as a research project no. N202 088 32/2030, N N202 1349 33 and NN202288338. We appreciate (A.S, M.W.) the financial support of the Project Based Personal Exchange Program with DAAD/MNiSW in years 2007–2008. We are grateful Dr. M. Richter from IFW Dresden for valuable discussions.

**Figure captions**

Fig.1. X-ray photoemission spectrum: a) valence band , b) U *4f* lines, c) in broad energy range (Note: U NOO, Si LMM etc. are Auger Lines).

Fig.2. Left panel: comparison of total DOS of ferromagnetic UCu$_2$Si$_2$ calculated with different approaches. (a) FPLO (LSDA) (differences between results obtained with LmtART, FPLO and Wien2k are negligible); (b) FPLO (LSDA+OP), (c) Wien2k (GGA+OP), (d) Wien2k (GGA+*U*) for $U_{5f}$=3eV, $J_{5f}$=0.5 eV. Right panel: DOS of ferromagnetic UCu$_2$Si$_2$ calculated with the Wien2k (GGA+*U*) method ($U_{5f}$=3eV, $J_{5f}$=0.5 eV). The total DOS (upper panel) and the U-, Cu-, and Si-site contributions (lower panels) are presented.

Fig.3. a) Comparison of the measured X-ray photoemission spectrum with related calculated data obtained with FPLO and Wien2k without spin polarization; b) The low binding energy range is shown for clarity. The computed contributions from U (mainly *5f* band) and Cu (mainly *3d* band) are also plotted; c) The same as b), but now using GGA+*U* with the following parameters: $U_{5f}$ = 1 eV together with three values of $U_{3d}$ = 2, 2.5, and 3.18 eV, as well as $U_{5f}$ = 1.5 eV and $U_{3d}$ = 2.8 eV.

Fig. 4. The dependence of the calculated magnetic moments on the uranium atoms in UCu$_2$Si$_2$ on the Coulomb repulsion *U* parameter taken in the GGA+*U* approach (Wien2k, see Table 2). For comparison also LSDA/GGA (ASW [8], Wien2k, FPLO, and LmtART) and LSDA/GGA+OP (ASW [9], Wien2k, and FPLO) values are drawn in the right-hand column (see Table 2). Experimental results of Refs. 2, 4 and 6 are marked by horizontal lines. Note that in Refs. [4, 6] the total magnetic moment per formula unit was reported. We find, however, that the magnetic moments on the Cu and Si sites can be neglected, see Table 2.

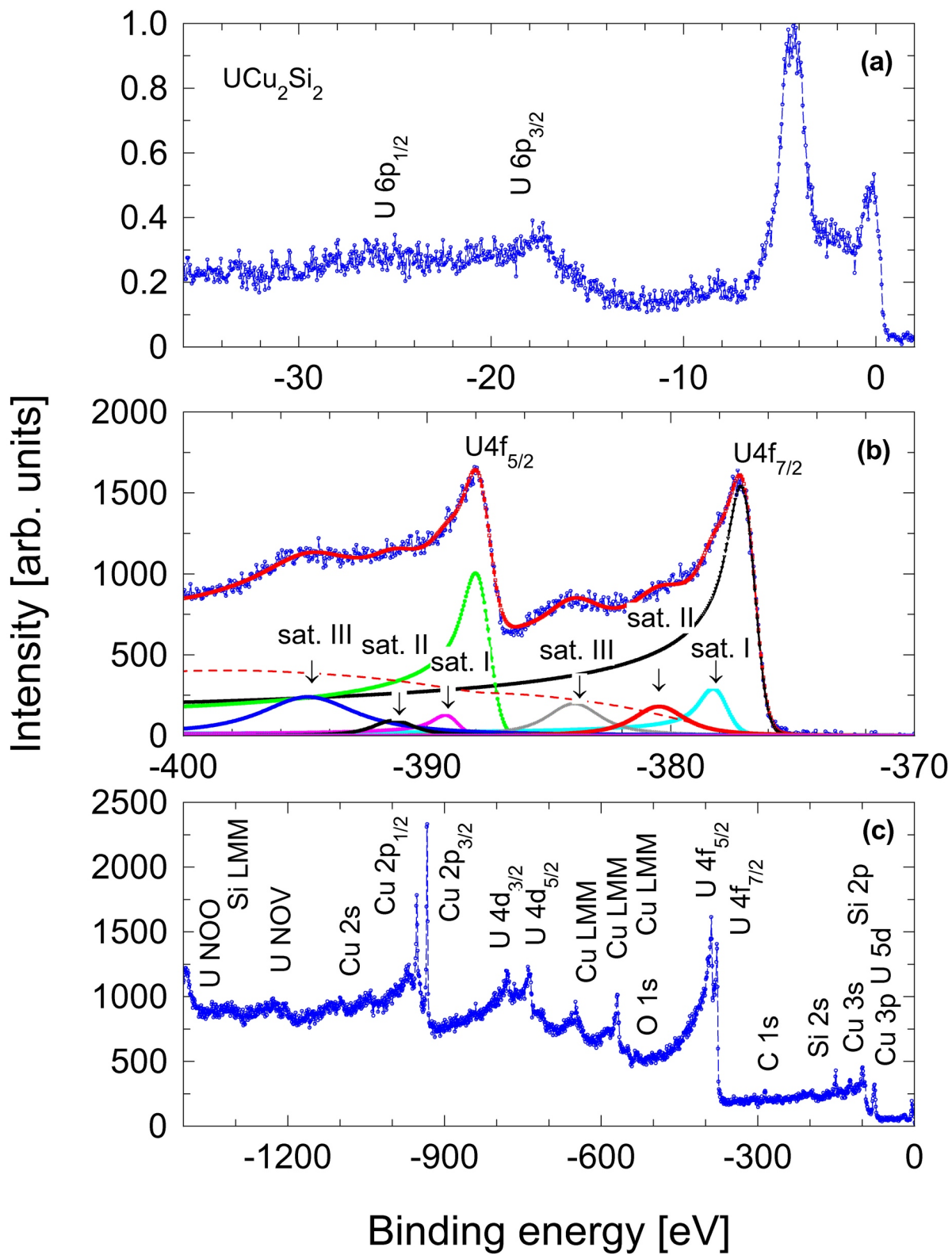

Fig.1

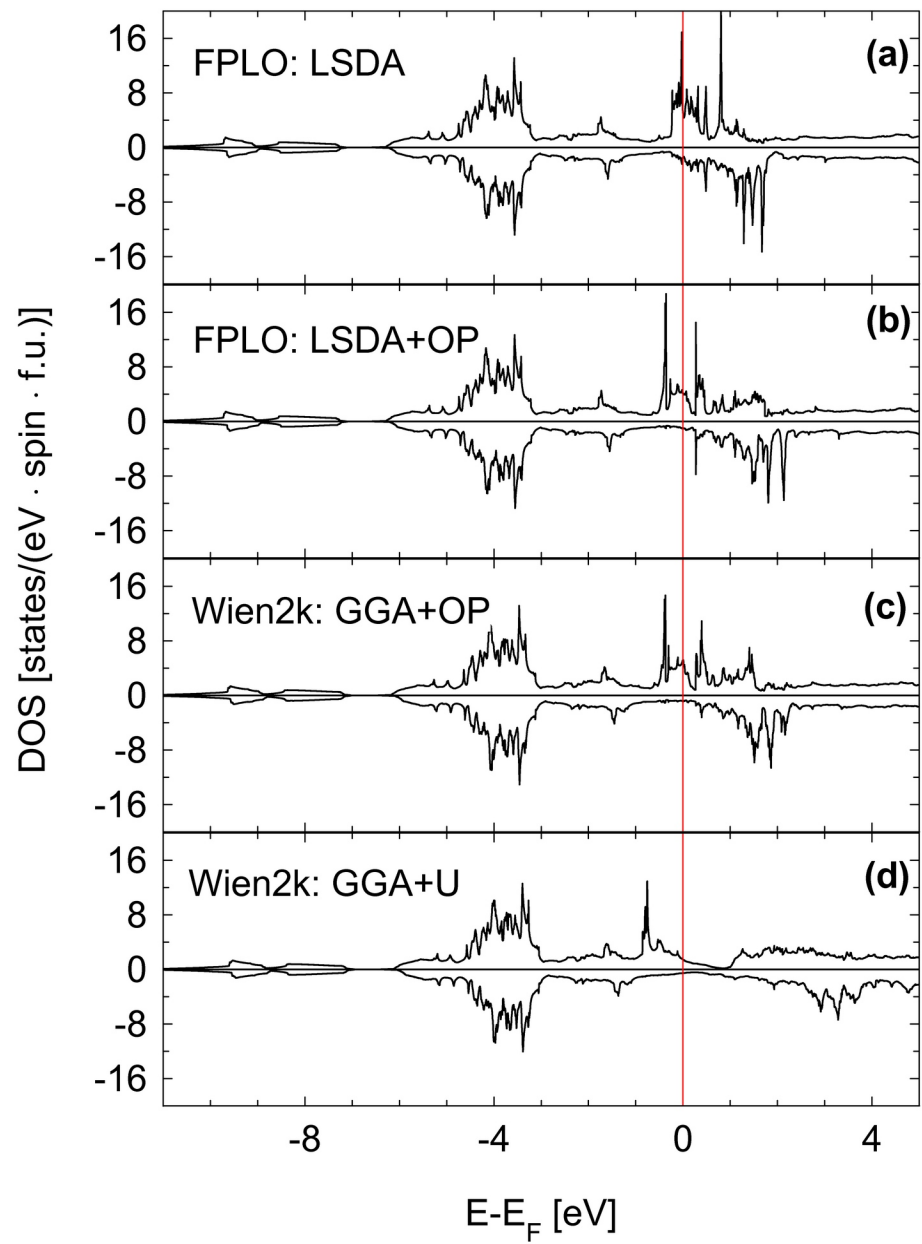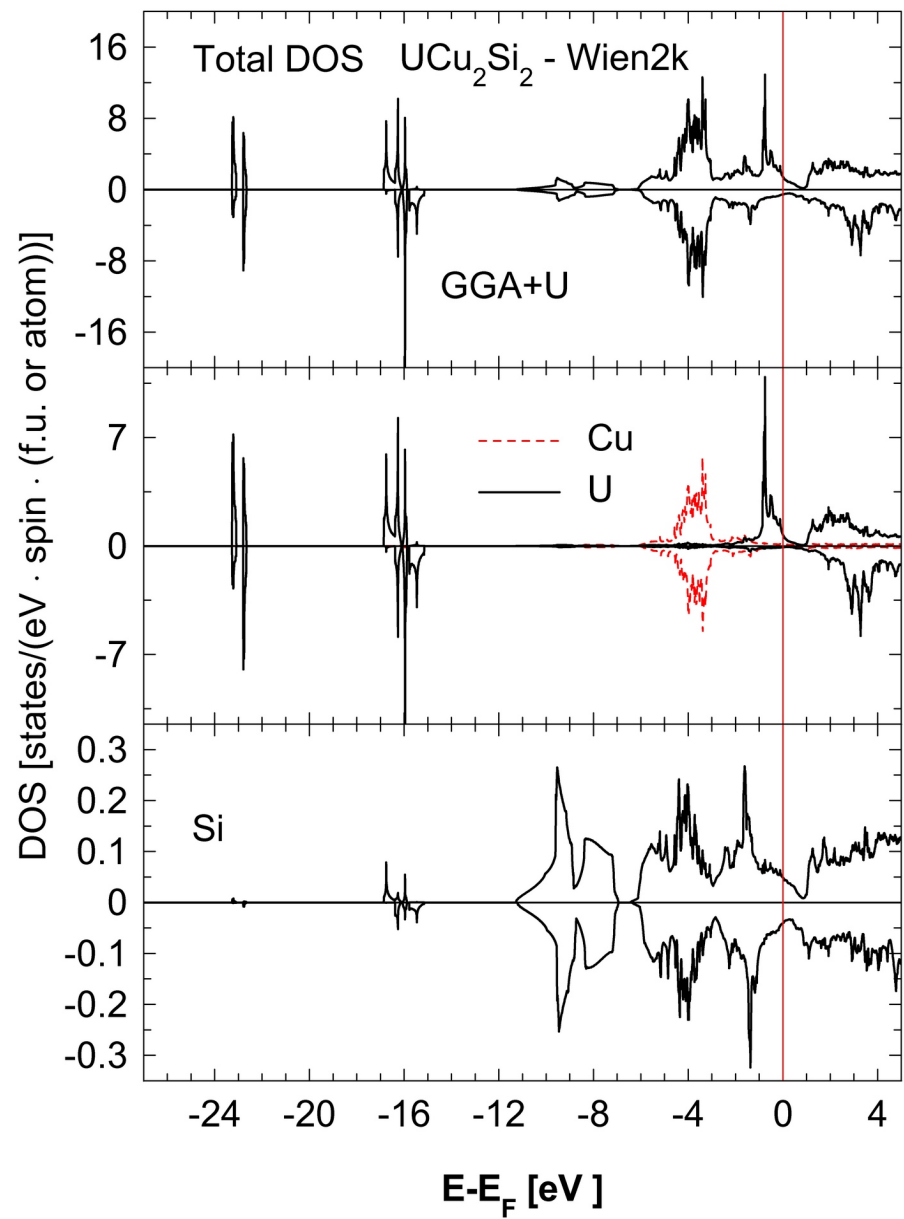

**Fig2**

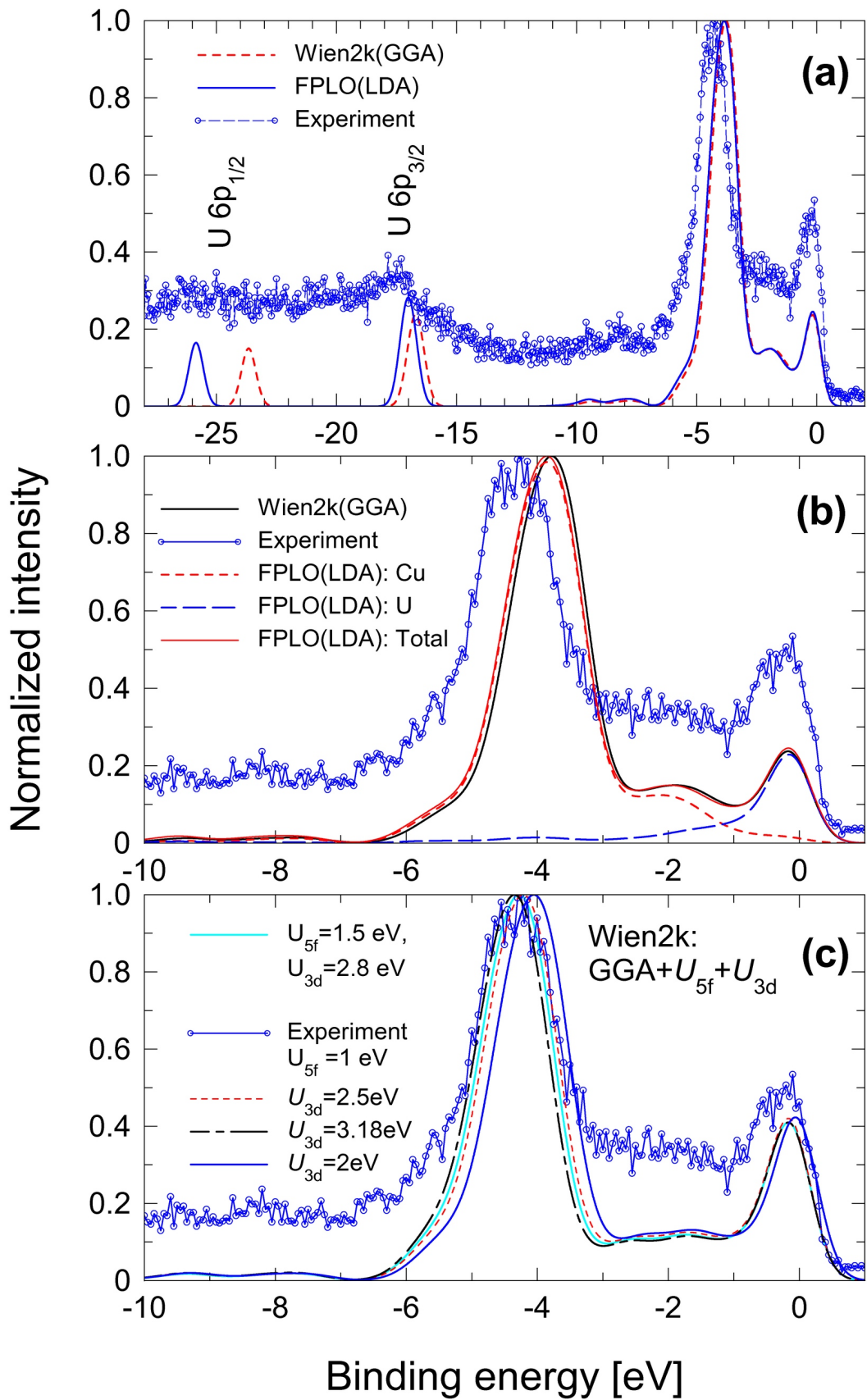

Fig. 3

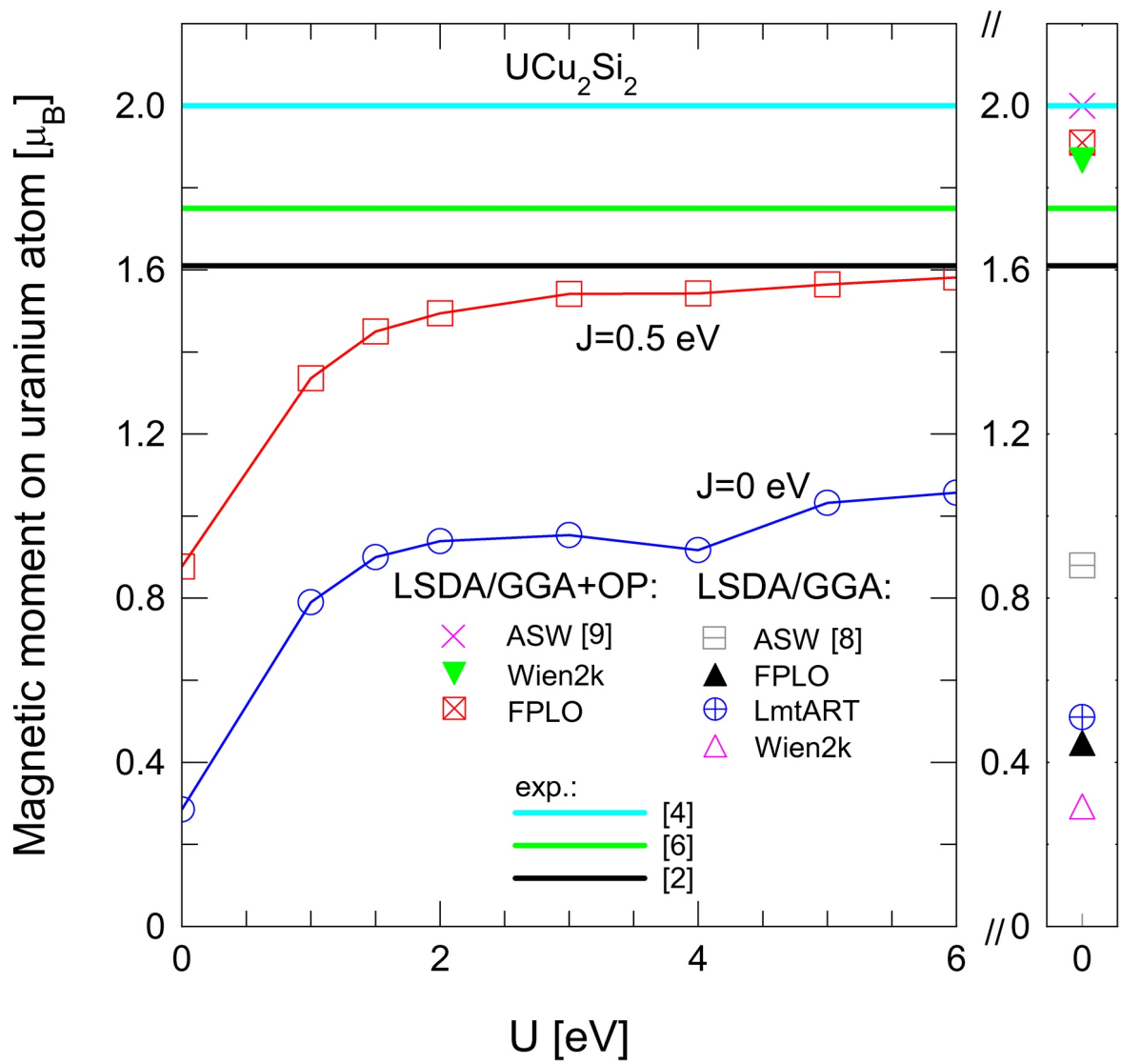

Fig4

**Table 1**

Spin projected densities of electronic states (DOS) at the Fermi level: total, [states/(eV f.u. spin)] and site projected on U atom [states/(eV atom spin)]. In the parenthesis contribution of *5f* electrons is presented. Signs ↑, ↓ denote spin directions. Calculated values of the Sommerfeld coefficient $\gamma_0$ [mJ/mol K$^2$]. Collected values were obtained using different methods of calculations:

I: FPLO(LSDA+OP); II: Wien2k(GGA+OP); III: FPLO(LSDA); IV: Wien2k(GGA);
V: Wien2k(GGA+*U*)- ($U_{5f}$ =3 eV; *J*=0.5 eV); VI: LmtART(GGA).

| Type of DOS/($\gamma_0$) Atom (position) | spin | Total and site-projected DOS | | | | | |
|---|---|---|---|---|---|---|---|
| | | I | II | III | IV | V | VI |
| U(2b) | ↑ | 3.49(3.21) | 3.83(3.75) | 4.15(3.91) | 4.35(4.23) | 0.75(0.65) | 4.68(4.40) |
| | ↓ | 0.54(0.44) | 0.29(0.22) | 0.98(0.85) | 0.69(0.64) | 0.08(0.03) | 0.94(0.82) |
| Total | ↑ | 4.19 | 4.97 | 4.78 | 5.74 | 1.46 | 5.22 |
| | ↓ | 0.94 | 0.91 | 1.42 | 1.24 | 0.54 | 1.20 |
| $\gamma_0$ | | 12.1 | 13.9 | 14.6 | 16.5 | 4.7 | 15.1 |

**Table 2**

Experimental and calculated magnetic moments for $UCu_2Si_2$ using different methods of calculations. Parameters used in the Wien2k LSDA+$U$ calculations: *: $U_{5f}$ = 1 eV; $J$ = 0.5 eV; **: $U_{5f}$ = 3 eV; $J$ = 0.5 eV. In the case of Refs. [4, 6], the total magnetic moment per formula unit was reported.

| Atom (position) | | m [$\mu_B$/atom] | | |
|---|---|---|---|---|
| | | spin | orbital | total |
| U(2b) | [6] | | | 1.75 |
| | Experiment [2] | | | 1.61 |
| | [4] | | | 2.00 |
| | LSDA: ASW [8] | -2.21 | 3.09 | 0.88 |
| | FPLO | -1.82 | 2.26 | 0.44 |
| | GGA: LmtART | -2.09 | 2.60 | 0.51 |
| | Wien2k | -1.81 | 2.09 | 0.29 |
| | GGA+$U$: Wien2k* | -2.12 | 3.46 | 1.34 |
| | Wien2k** | -2.42 | 3.96 | 1.54 |
| | LSDA+OP: ASW [9] | -2.80 | 4.80 | 2.00 |
| | FPLO | -2.17 | 4.08 | 1.91 |
| | GGA+OP: Wien2k | -2.01 | 3.88 | 1.87 |
| Cu(4d) | FPLO(LSDA+OP) | -0.02 | -0.01 | -0.03 |
| Si(4e) | FPLO(LSDA+OP) | -0.02 | 0.01 | -0.01 |